# Self-locking non-volatile coding metasurfaces via origami-based mechanical bits

Ding Zhang[1,2], Peng Tang[1,2], Liqiao Jing[1,2]*, Xincheng Yao[1,2], Bo Zhou[1,2], Enzong Wu[1,2], Ying Li[1,2],

Evgueni Filipov[3,4]*, Hongsheng Chen[1,2]*, and Zuojia Wang[1,2]*

[1] *Zhejiang Key Laboratory of Intelligent Electromagnetic Control and Advanced Electronic Integration, ZJU-Hangzhou Global Scientific and Technological Innovation Center, College of Information Science and Electronic Engineering, Zhejiang University, Hangzhou 310027, China.*

[2] *International Joint Innovation Center, The Electromagnetics Academy at Zhejiang University, College of Information Science and Electronic Engineering, Zhejiang University, Hangzhou 310027, China.*

[3] *Deployable and Reconfigurable Structures Laboratory, Department of Civil and Environmental Engineering, University of Michigan, Ann Arbor, Michigan, USA.*

[4] *Deployable and Reconfigurable Structures Laboratory, Department of Mechanical Engineering, University of Michigan, Ann Arbor, Michigan, USA*

*Corresponding authors. E-mail: liqiaoj@zju.edu.cn (L. Jing); filipov@umich.edu (E. Filipov); hansomchen@zju.edu.cn (H. Chen); zuojiawang@zju.edu.cn (Z. Wang);

**Abstract:** Digital coding metasurfaces have revolutionized electromagnetic (EM) manipulation, yet typical tunable approaches based on active components suffer from the "volatility" bottleneck. While mechanical modulation provides a potential solution, current implementations generally lack inherent state-locking capability, rendering them vulnerable to environmental disturbances and actuation errors. Inspired by the concept of mechanical bits (MBs), this paper presents a self-locking non-volatile coding metasurface platform enabled by Kresling origami-based MBs, where the continuous mechanical deformation of individual meta-atoms is discretized into robust binary geometric states protected by intrinsic energy barriers. The bistable states are strictly mapped to 1-bit EM coding phases via tailored metallic patterns integrated onto a multimaterial 3D printed Kresling origami array. Building upon this concept, both transmission- and reflection-type prototypes are proposed and experimentally demonstrated, exhibiting exceptional wavefront manipulation capabilities through near-field holographic imaging and far-field beam steering. In addition, the lightweight origami unit (1.5 g) exhibits an exceptional load-bearing capacity, supporting over 100 times its own weight. These results bridge mechanical logic with EM information processing, establishing a universal physical paradigm for constructing low-power, highly robust coding metasurfaces resilient to extreme environments.

## 1. Introduction

The paradigm shift from static structures to digital coding metasurfaces has revolutionized electromagnetic (EM) wave manipulation, driving breakthroughs in next-generation wireless communications[1,2]. By switching the on/off states or modulating the capacitance of active components (e.g., PIN diodes, varactors) integrated into independent meta-atoms, electrically tunable coding metasurfaces exhibit fast response speeds and sophisticated wavefront tailoring, facilitating various innovative functionalities, including adaptive beamforming[3], multiplexed vortex beams[4], reconfigurable meta-holograms[5], and programmable diffractive neural networks[6,7]. Recently, the temporal domain has been explored as an additional degree of freedom for coding metasurfaces[8], which has further broadened the scope of applications, such as Doppler cloaks[9], nonreciprocal antennas[10], and sideband-free radiation[11,12]. Despite the success of electrically tunable coding metasurfaces in digitizing EM responses, the requirement for complex biasing networks and external continuous power sources to maintain the coding patterns introduces a "volatility" bottleneck[13–16] and restricts their deployment in energy-constrained scenarios[17,18].

The mechanical modulation has recently emerged as a powerful framework for the dynamic control of EM waves[19–21], which leverage physical structural changes to tune EM properties, offering a promising avenue for applications requiring non-volatile memory and environmental robustness. Approaches based on stretchable substrates utilize the elastic deformation of elastomers (e.g., PDMS) to continuously modulate the lattice periodicity or inter-particle spacing of meta-atoms[22–24]. More sophisticated geometric transformations, such as origami- or kirigami-inspired designs, exploit geometric folding and twisting to reconfigure structural symmetries, enabling the dynamic manipulation of chirality and EM wavefronts[25–29]. Furthermore, recent advances have introduced motorized meta-atoms, where the independent rotation of local elements facilitates effective phase shifting, demonstrating remarkable tunability[30–33]. However, most mechanically reconfigurable metasurfaces lack a state-locking capability, rendering EM responses highly susceptible to environmental disturbances and actuation errors, with the attendant risk that slight structural damage could lead to a severe decline in signal fidelity[34]. Therefore, the shift from continuous physical deformation to discrete state switching with intrinsic stability is crucial for reliable information processing under long-term operation[35,36].

Multistable mechanical structures, characterized by their discrete stable states and inherent energy barriers, provide a potential method for addressing the stability challenges in reconfigurable systems[37,38]. By providing a passive mechanical threshold, the energy barriers can effectively filter out minor fluctuations, firmly locking the system into pre-defined geometric configurations without the need for continuous energy input. Owing to these non-volatile state-retention capabilities, such architectures facilitate the construction of mechanical bits (MBs)[39,40]. Recent developments have demonstrated their potential across diverse functions, including mechanical logic gates[41], data encryption[42], and mechanical neural networks[43,44]. In the realm of EM metasurfaces, a double-broadband absorbing grid enabled by bistable inflatable origami actuators has been proposed, allowing complementary absorption bands to be switched through structural reconfiguration[45]. A thin multifunctional metamaterial based on bistable curved beams has been developed, which supports programmable broadband absorption by binary-state encoding of unit cells[46]. More recently, a kirigami-origami hybrid metamaterial has been proposed to realize multiband and ultra-wideband switchable microwave absorption under large 3D deformations[47]. However, current multistable EM designs focused primarily on demonstrating macroscopic impedance matching (e.g., switchable absorbers), lacking independent, pixel-level phase manipulation required for information coding, thereby constraining their versatility for sophisticated spatial wavefront engineering.

In this work, we propose a self-locking non-volatile coding metasurface platform, consisting of a Kresling origami array, where the bistable states of individual units are strictly mapped to 1-bit EM coding phases. By leveraging the intrinsic bistability of the Kresling structure, continuous mechanical deformation of individual meta-atoms is discretized into two stable configurations protected by physical energy barriers, establishing a MB mechanism that effectively filters out environmental disturbances below the switching threshold. Building upon this concept, we design two distinct prototypes, a MB-enabled transmission coding metasurface (MB-TCM) and its reflection counterpart (MB-RCM) by integrating tailored metallic patterns onto multimaterial 3D printed Kresling units. Across their operating bandwidths, both prototypes achieve effective 1-bit phase reconfiguration (160°–200°) and experimentally demonstrate dynamic holographic imaging and beam steering, respectively, showcasing the system's exceptional wavefront engineering capacity. In addition, the Kresling unit is lightweight (1.5 g) yet highly stable, maintaining the expanded state under loads exceeding 100 times its own weight. The proposed architecture and method bridge the gap between mechanical logic and EM information, opening new pathways for low-power, high-robustness, and non-volatile intelligent communication systems tailored for deep-space exploration and other extreme environments.

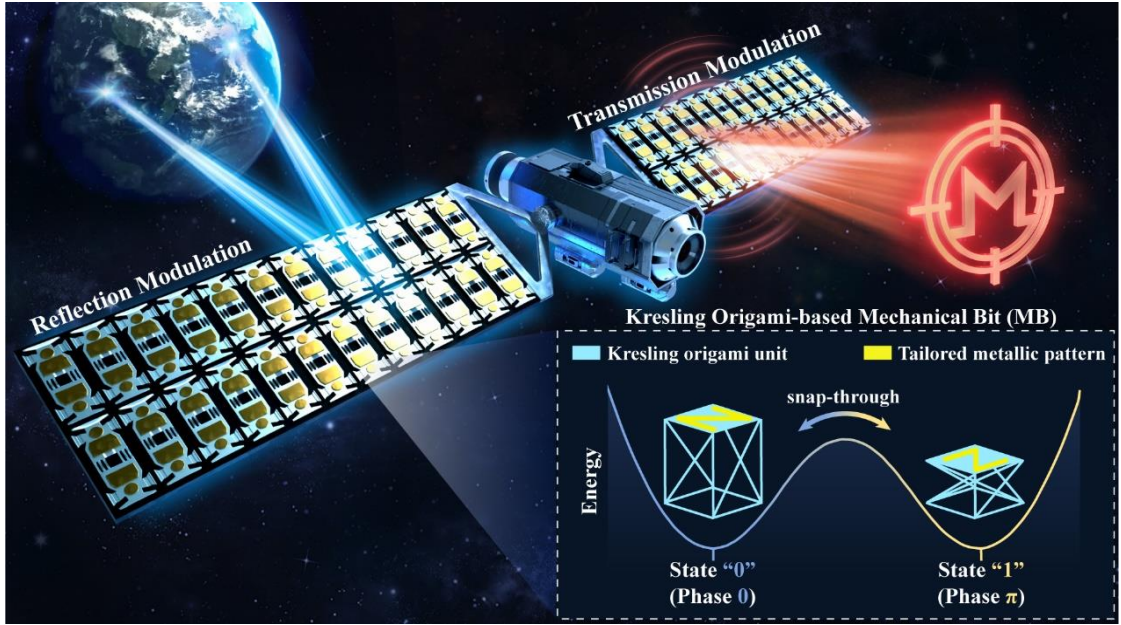

**Figure 1.** Conceptual design and application scenario of the proposed mechanically coding metasurfaces. The bistable states of individual Kresling meta-atom are strictly mapped to 1-bit EM coding phases. The mechanical-bit-enabled transmission coding metasurface (MB-TCM) and its reflection counterpart (MB-RCM) are realized by integrating tailored metallic patterns onto multimaterial 3D printed Kresling units, demonstrating dynamic holographic imaging and beam steering, respectively.

## 2. Results

### 2.1. Conceptual framework of self-locking non-volatile coding metasurfaces via Kresling origami-based mechanical bits

The schematic of the self-locking non-volatile coding metasurface platform via Kresling origami-based mechanical bits (MBs) is illustrated in **Figure 1**. Each MB integrates two key components to achieve mechanical reconfiguration and electromagnetic (EM) phase switching: Kresling origami units and metallic patterns. The bistability of the Kresling structure establishes an MB mechanism with state-locking capabilities, which can digitize continuous external actuation forces into two stable geometric states (state "0" and state "1") protected by physical energy barriers, thereby endowing the system with inherent robustness against environmental perturbations. Based on this mechanical platform, two types of robust non-volatile digital coding metasurfaces that employ different degrees of freedom to reconfigure the EM response are designed. The meta-atom of the MB-TCM is a "grating-split ring resonator (SRR)-grating" sandwich structure. It achieves a 1-bit transmission phase flip by driving the rotation of the middle-layer SRR within a polarization-selective Fabry–Pérot cavity formed by orthogonal gratings, enabling applications such as holographic imaging. Moreover, the MB-RCM operates by leveraging the axial height variation between two stable states of the Kresling unit, employing a Jerusalem Cross structure with four-fold (C4) rotational symmetry as the meta-atom to decouple the effects of rotation. To achieve stable reflection-phase switching for beam steering, this design adjusts the cavity's electrical length by varying the physical separation between the top resonator and the bottom ground plane. As a universal physical interface, the MB successfully translates robust mechanical bistability into programmable EM wavefronts through the strategy of digitizing mechanical deformation, expanding the potential applications.

**2.2. Geometry and mechanical properties of the Kresling structure**

The bistability of the proposed metasurface meta-atom stems from the Kresling origami topology. Guided by the geometric relationships depicted in the schematic (**Figure S1**, Supporting Information), the specific configuration of the designed Kresling pattern is illustrated in **Figure 2a**. In the deployed state (state "0"), the structure forms a cuboid with total height $h$ and square top/bottom surfaces with side length $a$. These geometric parameters are optimized to satisfy the "critical design"[48]:

$$a/h = \sqrt{3}/2 \tag{1}$$

Under this configuration, the Kresling structure exhibits unique kinematic characteristics, whereby the transition to the folded state (state "1") involves a 50% reduction in axial height and a 90° relative rotation between the basal surfaces. Such a prescribed geometric transformation is instrumental in engineering, as it provides the physical foundation for the system's intrinsic mechanical bistability.

As a typical non-rigid origami, the Kresling pattern undergoes mandatory panel deformation during folding, giving rise to its signature nonlinear mechanical behavior[49,50]. To characterize its bistability, a truss model that neglects the panel deformation is employed, as shown in **Figure S2a**. In this model, the strain energy $U$ is primarily stored within the elastic crease lines:

$$U = \frac{n}{2} k (|L'_{AC} - L_{AC}|^2 - |L'_{BC} - L_{BC}|^2) \tag{2}$$

where $n$ represents the number of the base polygon sides, $k$ is the elastic constant of the linear truss cell, and $|L'_{AC} - L_{AC}|$ and $|L'_{BC} - L_{BC}|$ denote the deformation of the crease lines $AC$ and $BC$. To validate the bistable characteristics, finite element method (FEM) simulations are performed. Here, with $a$ = 17.5 mm and $h$ = 20 mm fixed, which approximately satisfies the critical design, the elastic modulus of the basal members is set to $E_b$ = 2200 MPa, and the elastic modulus of the side members $E_s$ is varied. The bistable mechanics of the Kresling unit are collectively characterized through energy- and force-displacement curves as illustrated in **Figures 2b** and **2c**, indicating that enhancing side-member stiffness, which corresponds to a lower $E_b/E_s$ ratio, substantially elevates both the energy barrier and the critical buckling force required for snap-through switching. Crucially, while stiffness dictates the magnitude of the switching threshold, the coordinates of the energy minima and zero-force crossings remain remarkably invariant. This consistency confirms that the stable equilibrium states, defined by a near-50% height reduction, are strictly governed by geometric topology rather than material elasticity. Furthermore, the near-equivalent energy minima signify comparable barriers for bidirectional switching, ensuring balanced stability. These characteristics provide essential theoretical guidance for fabricating high-fidelity, non-volatile coding metasurfaces with excellent mechanical performance.

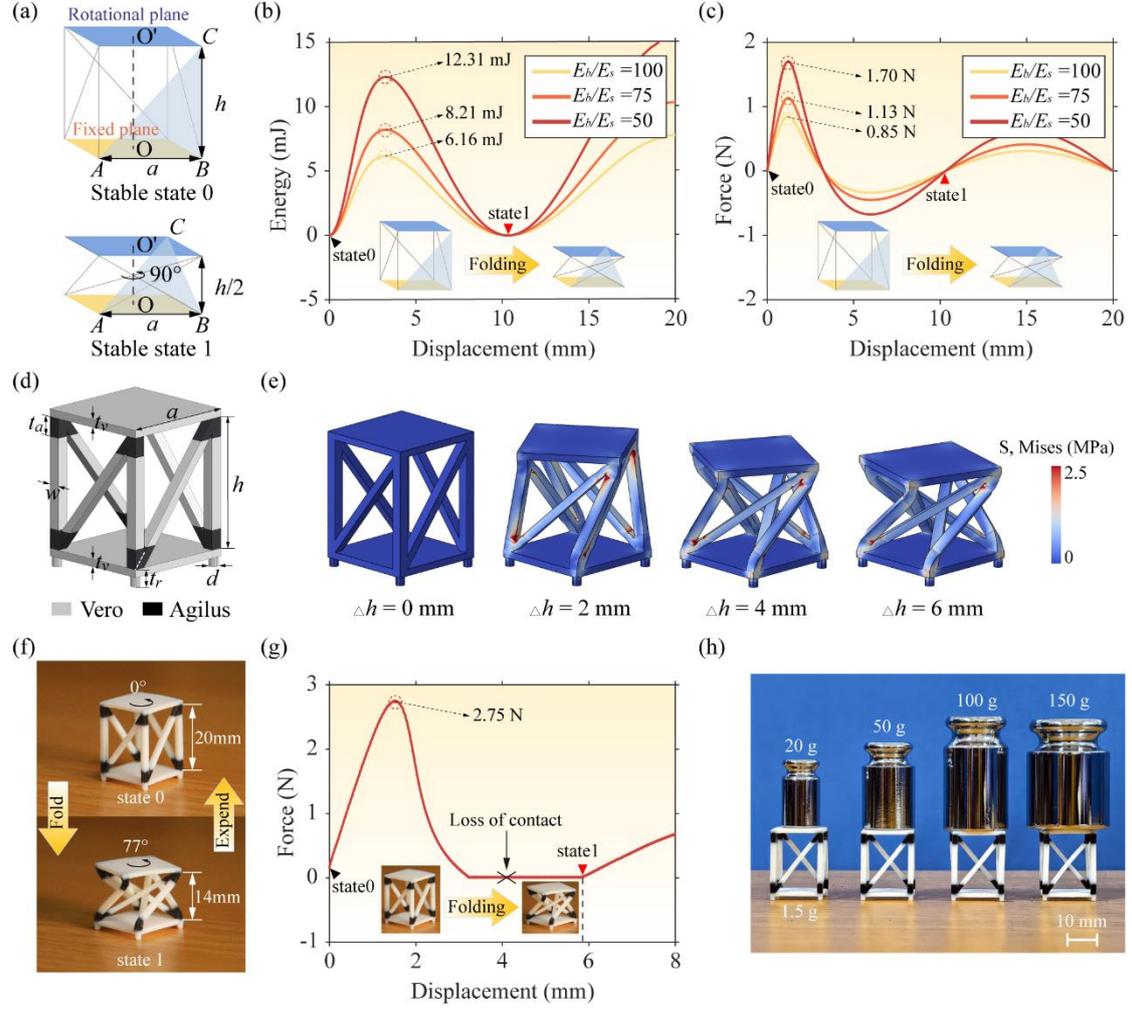

**Figure 2.** Geometric configuration and mechanical characterization of the Kresling unit. (a) Geometric illustration of the Kresling unit under critical design conditions. Simulated (b) energy-displacement and (c) force-displacement profiles of the truss model during the transition from state "0" to state "1" across various elastic modulus ratios $E_b/E_s$. (d) Detailed structural design and geometric parameters of the multimaterial 3D printed model. The white and black regions correspond to rigid (Vero) and flexible (Agilus) materials, respectively. (e) Simulated Mises equivalent stress distribution at different compression stages. (f) Photographs of the fabricated Kresling unit in the state "0" and state "1". (g) Experimentally measured force-displacement response curve of the Kresling unit. (h) Load-bearing test demonstrating the high structural stability of the unit in the expanded state.

To ensure reliable fabrication, the Kresling model was prototyped using multimaterial 3D printing, and its detailed dimensions are depicted in **Figure 2d**, where the white and black regions denote the Vero resin (rigid rods) and Agilus resin (flexible joints), respectively. Additionally, the cylindrical protrusions at the base corners facilitate assembly with the dielectric substrate. The optimized geometric parameters are: $a = 17.5$, $h = 20$, $t_v = 1.2$, $t_a = 3$, $w = 1.5$, $t_r = 1.5$, and $d = 1.5$, in millimeter. A 3D homogeneous solid model for the mechanical simulation is employed, as shown in **Figure S2b**. The Mises equivalent stress distribution diagrams of the simulation model, as illustrated in **Figure 2e**, reveal that stress is predominantly concentrated at the flexible joints, and mechanical interference in the sidewalls occurs when the compression displacement reaches approximately 6 mm. **Figure 2f** presents the fabricated Kresling unit, showing that the actual height difference between the two states is 6 mm, accompanied by a 77° relative rotation in state "1". The discrepancy between the theoretical predictions and measured results is attributed to the actual physical dimensions of the printed members, which violate the ideal zero-thickness assumption. To demonstrate the bistability experimentally, **Figure 2g** presents the measured force-displacement curve, which can be divided into three distinct segments. In the first segment ($\Delta h$ < 3.2 mm), the Kresling structure undergoes elastic compression, with the reaction force rising to a peak of 2.75 N at $\Delta h$ = 1.8 mm before dropping to 0 N at $\Delta h$ = 3.2 mm. In the second segment ($\Delta h$ ≈ 3.2–5.9 mm), the Kresling structure undergoes snap-through buckling, transitioning from state "0" to state "1". During this phase, the structure loses contact with the force sensor, resulting in a measured force of 0 N. In the third segment ($\Delta h$ > 5.9 mm), the sensor re-engages with the structure, and the reaction force rises again as the structure enters the densification regime. Furthermore, the Kresling unit features a lightweight architecture (1.5 g) with exceptional load-bearing capacity, supporting 100 times its own weight in the deployed state without structural deformation, as illustrated in **Figure 2h**. Additionally, **Figure S3** reveals robust performance retention over 200 compression-torsion cycles in the durability tests, confirming the unit's exceptional fatigue resistance and long-term stability.

**2.3. Electromagnetic reconfigurability of the meta-atoms**

The inherent bistability of the Kresling structure serves as a robust physical basis for constructing a reconfigurable EM platform. To verify the versatility of this platform, full wave simulations in commercial software CST Microwave Studio are performed with unit boundaries along *x*- and *y*-directions. The structural schematics of the MB-TCM and MB-RCM meta-atoms are illustrated in **Figures 3a** and **3b**, respectively, both with a lattice period of 22 mm. The optimized geometric parameters and surface current analyses for these meta-atoms are shown in **Figure S4** (Supplementary Information). As the resin rods are electrically small relative to the operating wavelength, their influence on EM wave scattering is minimal. Consequently, the rods were omitted from the simulation models to optimize computational efficiency. This simplification is validated in **Figure S5**, which demonstrates a negligible discrepancy for the meta-atoms between models with and without strut modeling.

For the MB-TCM meta-atom, the unit adopts a "grating-SRR-grating" sandwich configuration, in which the top and bottom surfaces consist of two mutually orthogonal metallic gratings, while the middle layer is an SRR attached to the rotatable surface of the Kresling structure. The orthogonal gratings act as a polarization filter effectively blocking direct transmission of co-polarized waves while forming a high-quality resonant cavity that reinforces the interaction between incident waves and the middle-layer SRR. The EM response of the MB-TCM meta-atom is plotted in **Figure 3c**. The unit maintains a high cross-polarized transmission coefficient ($|T_{xy}|$ > 0.75) in both stable states across the broad frequency range of 3.5–5 GHz. Once the unit transitions from state "0" to state "1", the SRR undergoes a precise geometric rotation. This rotation, combined with the cross-polarization coupling mechanism of the Fabry–Pérot cavity[51], triggers a near 180° transmission phase flip for the cross-polarized component. The phase difference remains remarkably stable within the range of 160°–200°, effectively fulfilling the requirement for robust 1-bit phase coding, despite slight rotational deviations from the ideal 90° dictated by the Pancharatnam-Berry (PB) phase principle[52]. This EM reconfiguration stems directly from harnessing the rotational degree of freedom.

For the MB-RCM meta-atom, in contrast to the rotation-based phase modulation used in the transmission unit, the reconfiguration strategy here shifts to exploiting the axial height variation of the Kresling structure, effectively decoupling the EM response from its inherent twisting motion. This decoupling is enabled by the Jerusalem Cross resonator with $C_4$ rotational symmetry, which ensures its EM response invariant to in-plane orientation[53]. Consequently, the phase switching depends solely on the vertical displacement between the resonator and the metallic ground plane. When the Kresling unit transitions from the deployed to the folded state, the cavity spacing undergoes an abrupt reduction, which directly alters the effective electrical length, shifting the resonant condition and producing the observed near 180° phase shift. The EM response of the MB-RCM meta-atom is plotted in **Figure 3d**, clearly showing a high co-polarized reflection coefficient ($|R_{xx}| > 0.85$) across 12.2–13 GHz in both stable states, accompanied by a stable phase difference of 160°–200°, successfully validating its effectiveness as a reflective 1-bit coding unit.

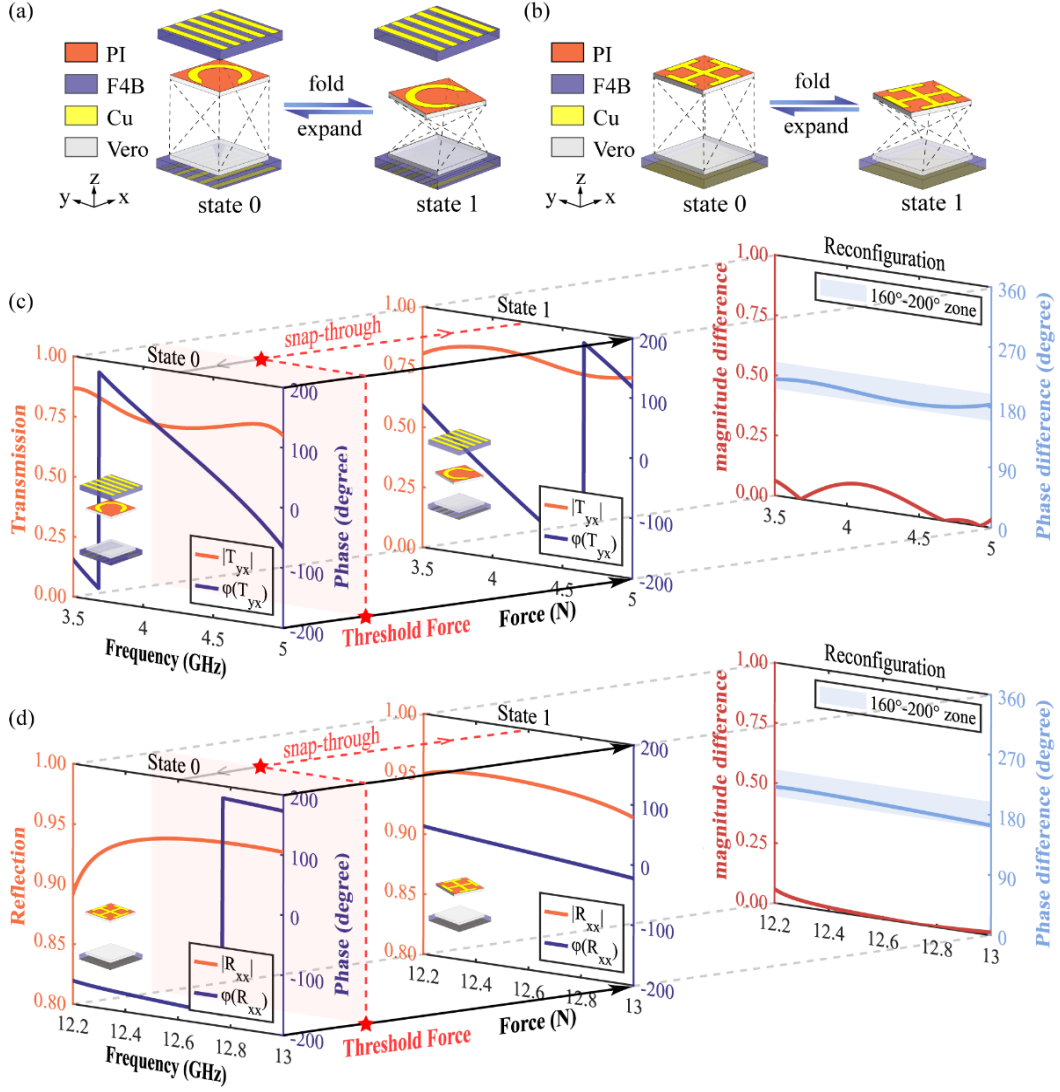

**Figure 3.** Structural schematics and reconfigurable characteristics of the MB-TCM and MB-RCM meta-atoms. (a) Schematic of the MB-TCM unit, featuring a "grating-SRR-grating" sandwich architecture designed for polarization-selective transmission. (b) Schematic of the MB-RCM unit, consisting of a Jerusalem cross and a metallic backplane. (c) Simulated EM response of the MB-TCM unit, demonstrating high cross-polarized transmission coefficient $|T_{xy}|$, and a stable phase difference between the two states. (d) Simulated EM response of the MB-RCM unit, showing high co-polarized reflection coefficient $|R_{xx}|$, and a stable phase difference between the two states.

Crucially, throughout EM reconfiguration, the Kresling unit functions as an MB with intrinsic memory. During the compression process, for instance, the structure exhibits a well-defined mechanical threshold, which means that any external force below this threshold triggers elastic recovery to state "0" and only forces exceeding it induce snap-through instability, locking the structure stably into state "1". This bistable snap-through behavior delivers two key advantages: zero static power consumption (energy is solely expended during switching) and inherent disturbance immunity (sub-threshold vibrations or contacts are mechanically filtered out). These features make the proposed approach significantly superior to conventional electrically tunable metasurfaces based on PIN diodes or varactors in terms of non-volatility and environmental robustness.

**2.4. Near-field holographic**

As a revolutionary imaging technique capable of recording and reconstructing EM amplitude and phase information, metasurface holography has demonstrated immense potential in information storage and image display. Analogous to artificial neural networks, metasurface holography can be conceptualized as a diffractive neural network (DNN) comprising an input layer, a single hidden layer, and an output layer[54]. This architecture enables efficient wavefront control and ultrafast information processing through light propagation. The densely packed unit cells within the metasurface act as neurons in the hidden layer. According to Huygens' principle and the Rayleigh–Sommerfeld diffraction equation, the field intensity at any position $\vec{r}^{l+1} = (x^{l+1}, y^{l+1}, z^{l+1})$ in the subsequent layer is determined by the collective contribution of all superatoms in the preceding layer, which can be expressed as:

$$E(\vec{r}^{l+1}) = \iint_{-\infty}^{\infty} \frac{z - z_i}{r^2} \left( \frac{1}{2\pi r} + \frac{1}{j\lambda} \right) exp\left( \frac{j2\pi r}{\lambda} \right) \cdot E(\vec{r}_i^l) \cdot T(\vec{r}_i^l) dxdy \quad (3)$$

where $\vec{r}_i^l = (x_i^l, y_i^l, z_i^l)$ denotes the position of the $i$-th neuron in the $l$-th layer, $\lambda$ is the incident wavelength, and $r = ((x^{l+1} - x_i^l)^2 + (y^{l+1} - y_i^l)^2 + (z^{l+1} - z_i^l)^2)^{1/2}$ represents the Euclidean distance between two neurons in adjacent layers. The complex transmission coefficient of each unit is denoted as $T(\vec{r}_i^l) = a_i^l(x_i^l, y_i^l, z_i^l) \cdot \exp(j\varphi_i^l(x_i^l, y_i^l, z_i^l))$, where $a_i^l(x_i^l, y_i^l, z_i^l)$ and $\varphi_i^l(x_i^l, y_i^l, z_i^l)$ represent the amplitude and phase, respectively. To simplify the calculation, the amplitude term can be treated as constant. Assuming the DNN consists of $M$ hidden layers, the energy intensity at the output plane is $s_k^{M+1} = |E_k^{M+1}|^2$, and the loss function is defined as the weighted mean square error (MSE) between the target field $g_k^{M+1}$ and the actual predicted field $s_k^{M+1}$:

$$f(\phi_i^l) = \frac{1}{K} \sum_k w_k (s_k^{M+1} - g_k^{M+1})^2 \quad (4)$$

In the DNN framework for our MB-TCM holography, the number of hidden layers is $M = 1$. $K$ represents the total number of sampling points in the output layer, and $w_k$ denotes the weight assigned to the $k$-th sampling point. By assigning higher weights to regions requiring field energy concentration, noisy areas can be effectively suppressed, thereby significantly enhancing the overall image quality.

To demonstrate the imaging capabilities of the MB-TCM experimentally, a prototype was fabricated and measured using the setup shown in **Figure 4a**. The prototype adopts the polarization-selective sandwich architecture designed for cross-polarization manipulation, with its top and bottom layers consisting of orthogonal metallic gratings, as presented in **Figure 4b** and **4c**. The internal region of the sample consists of a 16 × 16 array of unit cells, as depicted in **Figure 4d** and **4e**, with a total aperture size of 352 ×352 mm² (5.6λ × 5.6λ at 4.8 GHz). Each unit can independently switch states through mechanical folding and expanding, enabling pixel-level programming of the aperture phase. To validate the reconfigurability of the MB-TCM, the letters "Z", "J", and "U" are selected as target images shown in **Figure 4f**, corresponding to three distinct coding configurations. **Figure 4g** displays the 1-bit phase coding matrices optimized for an operating frequency of 4.8 GHz and a focal plane distance of 100 mm, using the modified simulated annealing (SA) algorithm, as detailed in **Text S5** (Supplementary Information). The optimization efficiency is further evidenced by the iterative decay of the loss function, as illustrated in **Figure S6** (Supplementary Information). The theoretical holograms calculated based on different phase distributions and Equation (3) are shown in **Figure 4h**, which clearly reconstruct the target characters. **Figure 4i** presents the simulation results obtained with the commercial software CST Microwave Studio. It is worth noting that the optimal imaging plane observed in the full-wave simulation is located approximately 65 mm from the lower surface of the MB-TCM, which is smaller than the theoretically preset distance of 100 mm, as shown in **Figure S7**. This discrepancy arises because the theoretical model approximates the metasurface as an ideal zero-thickness phase screen, neglecting the physical height of the Kresling units, as detailed in **Text S6**. The experimental results, as shown in **Figure 4j**, clearly reconstruct the contours of the three target letters. Furthermore, the broadband robustness of the imaging system is confirmed by the normalized holographic images at different frequencies, as shown in **Figure S8**. The high consistency among the theoretical, simulated, and experimental results strongly demonstrates the effectiveness of the proposed MB-TCM in near-field wavefront engineering.

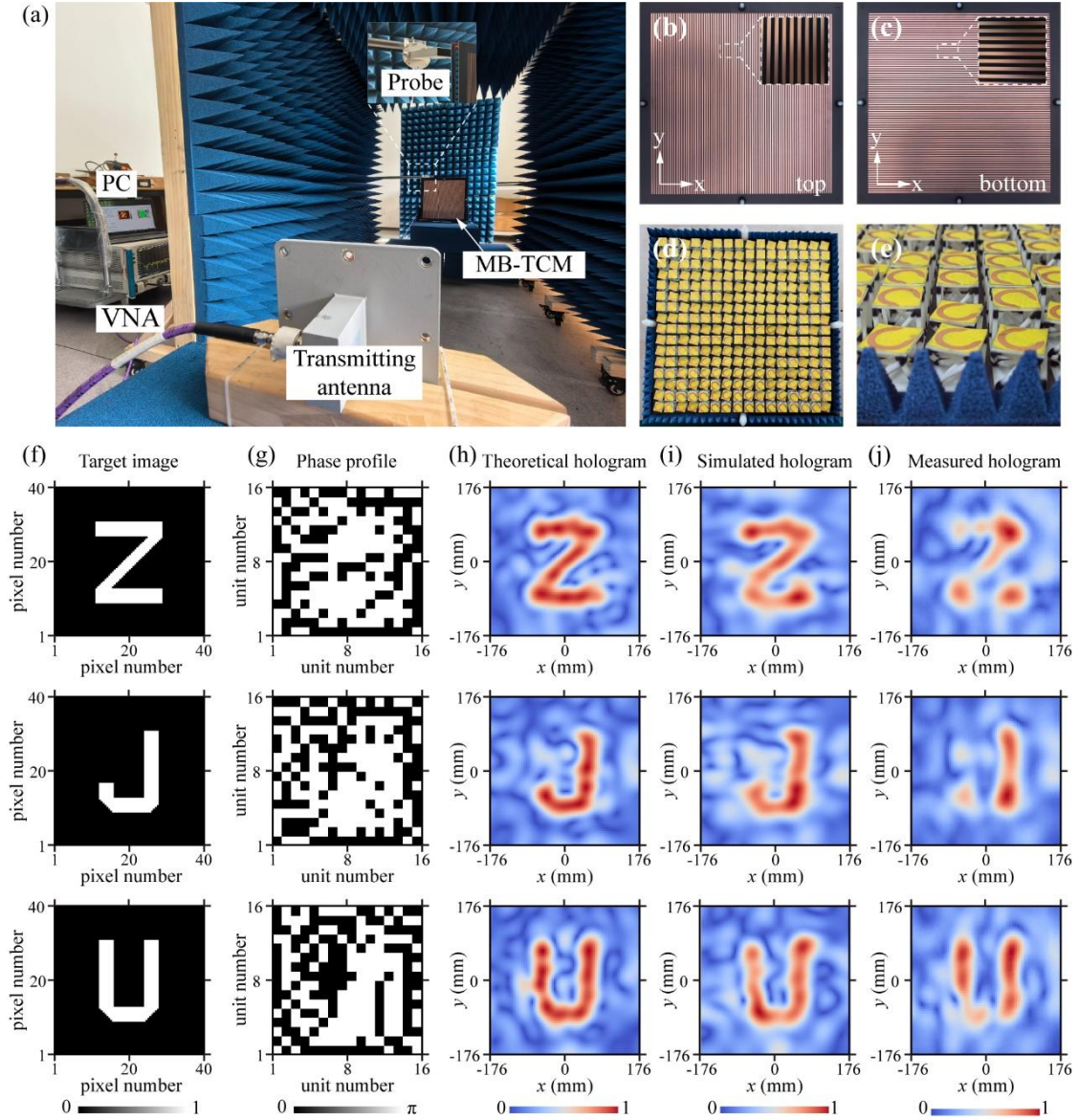

**Figure 4.** Near-field holographic imaging demonstration of the MB-TCM. (a) Photograph of the near-field scanning experimental setup. (b, c) Photographs of the fabricated MB-TCM prototype, displaying the top (b) and bottom (c) layers with orthogonal metallic grating structures. (d) Internal view of the array with the top grating layer removed to reveal the Kresling units. (e) Close-up view of individual unit cells in distinct coding states. (f) The three target holographic images (letters "Z", "J", "U"). (g) Corresponding optimized 1-bit phase coding matrices, where black and white pixels represent 0 and $\pi$ phase responses, respectively. (h–j) Comparison of the holographic imaging results at 4.8 GHz. Rows represent distinct letter patterns, while columns display (h) theoretical calculations based on the Rayleigh-Sommerfeld diffraction integral, (i) full-wave numerical simulations, and (j) experimental near-field scanning measurements.

**2.5. Far-field beam steering**

Based on the proposed mechanical reconfiguration methodology, the MB-RCM is further employed for beam steering. According to antenna array theory[55], the far-field scattering pattern of the metasurface can be derived by coherently superimposing the contributions from all individual unit cells. Under normal plane wave incidence, the far-field scattering function $F(\theta, \varphi)$ can be expressed as:

$$F(\theta,\varphi) = f_e(\theta,\varphi)\sum_{m=1}^{N}\sum_{n=1}^{N}\exp\{j[\varphi(m,n) + kD\sin\theta(m\cos\phi + n\sin\phi)]\} \tag{5}$$

where $\theta$ and $\varphi$ denote the elevation and azimuth angles in the spherical coordinate system, respectively. $f_e(\theta, \varphi)$ is the element factor, $(m, n)$ denotes the indices of the unit cell in the 2D array, corresponding to the $x$ and $y$ axis directions, $k$ is the free-space wavenumber, $D$ is the period of the metasurface unit, and $\varphi(m, n)$ denotes the coding phase distribution. For 1-bit coding, the value of $\varphi(m, n)$ is discretized to either 0 or $\pi$.

By digitally programming the spatial phase distribution $\varphi(m, n)$, the wavefront propagation direction of the scattered beam can be flexibly controlled. To validate this capability, four typical coding matrices are designed based on an 8 × 8 MB-RCM array as shown in **Figure 5a**: two uniform coding patterns ("0000..." and "1111...") and two periodic coding sequences ("0101..." and "0110..."). The corresponding simulated far-field radiation patterns are presented in **Figure 5b**. For the uniform coding configurations, all units exhibit in-phase reflection responses, rendering the MB-RCM electromagnetically equivalent to a perfect metallic plate. The incident EM wave undergoes specular reflection, with its energy highly concentrated in the normal direction. For the "01010101" coding sequence, a $\pi$ phase shift between adjacent rows is introduced, leading to beam splitting and generating a symmetric dual-beam scattering pattern. For the more complex "01100110" sequence, a quad-beam scattering pattern is further achieved. The fabricated MB-RCM prototype configured in the four distinct coding states is presented in **Figure 5c**. Benefiting from the independent mechanical bistability of the Kresling units, the array can be precisely configured and mechanically locked into any predefined geometric pattern without continuous power input.

The far-field measurement setup is illustrated in **Figure 5d**. A pair of horn antennas operating within the 11.9–18 GHz was connected to a vector network analyzer (VNA) to serve as the transmitter and receiver for co-polarized EM waves. The simulated and measured normalized far-field scattering patterns in the *xoz*-plane at 12.5 GHz are depicted in **Figure 5e**, showing high agreement in terms of main-lobe direction and beamwidth. Notably, when the receiving antenna is positioned close to the surface normal, it physically obstructs the EM waves from the transmitting horn, leading to a significant drop in the measured power within that angular range. These results confirm the reliable beam steering capability of the MB-RCM under mechanical coding control. Furthermore, to evaluate the spectral stability of the proposed platform, the far-field scattering patterns were measured across 1 GHz bandwidth. As illustrated in **Figure S9**, the radiation patterns for all four coding states maintain high pointing stability from 12.0 to 13.0 GHz, further validating the broadband effectiveness of the MB-RCM in diverse wavefront manipulation scenarios.

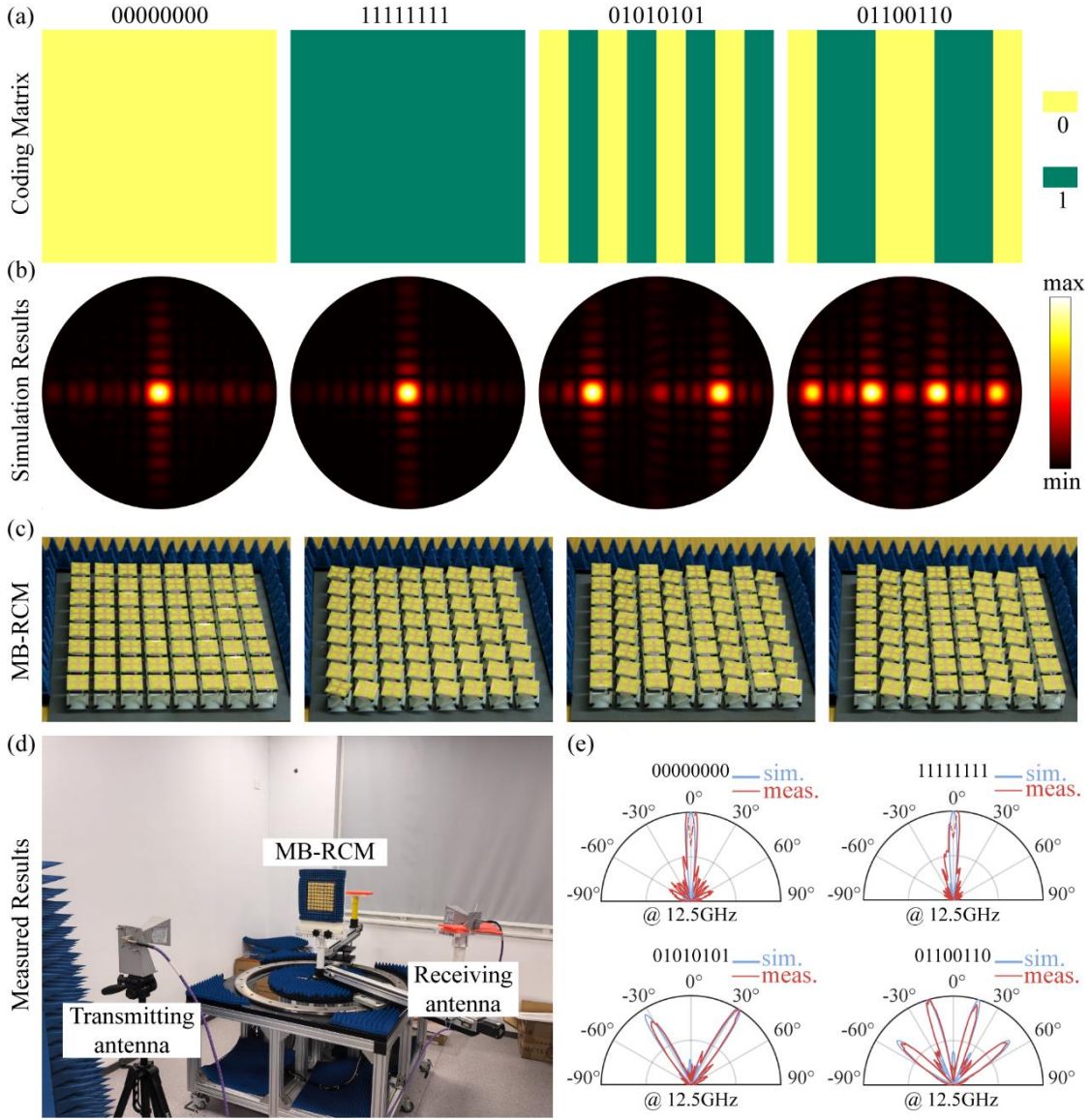

**Figure 5.** Far-field beam steering demonstration of the MB-RCM. (a) The four reconfigurable phase coding masks designed for the 8 × 8 array, corresponding to the "all-0" state ("0000..."), "all-1" state ("1111..."), dual-beam splitting state ("0101..."), and quad-beam splitting state ("0110..."). Yellow and green pixels represent the "0" and "1" states, respectively. (b) Simulated far-field radiation patterns at 12.5 GHz for the respective coding configurations above. (c) Photographs of the MB-RCM prototype under the respective coding states. (d) Photograph of the far-field measurement setup. (e) Comparison between simulated (blue lines) and measured (red lines) normalized far-field scattering patterns in the *xoz*-plane at 12.5 GHz. The dashed regions indicate the angular range where the receiving antenna physically obstructs the transmitting antenna, resulting in a measurement shadowing effect.

## 3. Discussion

In conclusion, a robust coding metasurface platform enabled by origami-based MB for non-volatile 1-bit phase modulation is proposed. By exploiting mechanical snap-through instability, continuous mechanical deformations can be discretized into robust binary geometric states. This mechanism confers unique advantages, including zero static power consumption and intrinsic error-correction capabilities, effectively circumventing the volatility constraints of electrically controlled devices and the stability limitations of conventional analog mechanical tuning. The bistable feature of the Kresling origami unit was systematically analyzed through both theoretical and numerical models. Experimental results further demonstrate that the lightweight sample fabricated by multimaterial 3D printing technology possesses exceptional structural repeatability and outstanding load-bearing capacity, capable of supporting loads exceeding 100 times its own weight. Based on the excellent physical foundation, the MB-TCM and MB-RCM meta-atoms achieve efficient EM responses, $|T_{xy}| > 0.75$ and $|R_{xx}| > 0.85$, respectively and a stable phase difference of 160°–200°. Functional prototype arrays were also fabricated and measured, successfully demonstrating exceptional wavefront manipulation capabilities at different frequencies by high-fidelity near-field holographic imaging and pointing-stability far-field beam steering. The high consistency observed among theoretical predictions, numerical simulations, and measurements fully validates the reliability of our design. The independent bistability of each unit is inherently compatible with future integration into array-scale actuation systems and a versatile paradigm is established deeply coupling mechanical logic with EM information processing. It provides a novel physical platform for constructing non-volatile, low-power, and environmentally robust programmable metasurfaces in energy-constrained environments, laying a solid foundation for future explorations into multi-bit coding, automated array addressing, and miniaturization.

## 4. Methods

**Mechanical simulation**: The mechanical numerical analysis model was established using the commercial Abaqus software. For the idealized truss model, the geometric vertices were defined as nodes connected by linear truss elements (T3D2). Fixed displacement constraints were applied to the nodes of the bottom square, while uniform vertical displacement was applied to the top nodes to drive the folding process. The cross-sectional area of all struts was set to 1 mm². The elastic modulus of the top and bottom struts was fixed at 2200 MPa, while the modulus of the side struts was varied (22, 29.3, and 44 MPa) to investigate the impact of sidewall flexibility on the bistable energy landscape. For the physical Kresling prototype, a 3D solid model was developed to evaluate stress distribution. The geometry was partitioned into distinct regions: the rods were modeled as a linear-elastic material ($E = 2200$ MPa, $v = 0.35$), while the soft joints were characterized using a hyperelastic Mooney-Rivlin constitutive model ($C_{10} = 0.4$, $C_{01} = 0.1$, $D_1 = 0$). Both models were solved using a static, general step.

**Electromagnetic simulation**: Full-wave simulations were conducted using CST Studio Suite to evaluate the EM performances of the MB-TCM and MB-RCM meta-units. The dielectric substrates were modeled as F4B high-frequency laminates (relative permittivity $\varepsilon_r = 2.65$ for MB-TCM and $\varepsilon_r = 2.2$ for MB-RCM). The polyimide film (0.05 mm thickness) was assigned a permittivity of $\varepsilon_r = 3$, and the copper layers were defined with an electrical conductivity $\sigma = 5.8 \times 10^7$ S/m. The Vero resin was modeled with $\varepsilon_r = 3.05$ and a loss tangent of $\tan\delta = 0.06$. Unit cell boundary conditions were applied along the $x$- and $y$-axes to simulate an infinite periodic array, while open (add space) boundary conditions were employed in the $z$-direction. For the holographic imaging and beam steering simulations, the metasurfaces were illuminated by plane waves, and field monitors were utilized to record the near-field distributions and far-field radiation patterns within the operating bands.

**Prototype fabrication**: The Kresling units were fabricated via PolyJet multimaterial 3D printing (Stratasys J850, Stratasys Ltd.). Each unit comprises rigid rods printed from VeroWhitePlus (Shore D hardness 85) and flexible joints printed from rubber-like Agilus material (Shore A hardness 50). The functional EM layers were fabricated using standard flexible printed circuit (FPC) techniques, consisting of polyimide/copper (0.1 mm/18 μm) films. The SRR and Jerusalem Cross patterns were obtained via photolithography and wet etching, then laser-cut into individual patches. These films were bonded to the Kresling units, which were subsequently assembled into perforated F4B dielectric substrates to form the final MB-TCM and MB-RCM arrays. The MB-TCM prototype is a 16 × 16 array (352 × 352 mm²), and the MB-RCM prototype is an 8 × 8 array (176 × 176 mm²).

**Mechanical experiments**: Quasi-static compression tests were conducted using a universal testing machine (Instron 34SC-1) equipped with a high-precision 50 N load cell (resolution 0.01 N). A displacement-controlled load was applied at a constant rate of 0.2 mm/s. To ensure stability during testing, the four cylindrical feet of the Kresling unit were inserted into a customized 3D printed PLA base fixture, which was secured to the compression platen. Detailed testing processes are provided in **Movie S1** (Supporting Information).

**Electromagnetic experiments**: Near-field holographic imaging was performed in an anechoic chamber. A standard horn antenna (3.94–5.99 GHz) served as the transmitter, placed 2 m away from the MB-TCM to generate an *x*-polarized quasi-plane wave. The imaging plane was located 65 mm behind the MB-TCM. A waveguide probe raster-scanned the 352×352 mm² area with a step size of 4.4 mm to map the *y*-polarized electric field component. Data acquisition was synchronized using a VNA.

For far-field beam steering, a bistatic measurement setup was employed using a pair of horn antennas (11.9–18 GHz) connected to a VNA. The transmitting horn was positioned 1.5 m from the MB-RCM. The prototype and the receiving horn were mounted on a rotary stage with a 1 m separation. The receiving antenna was rotated in the *xoz*-plane from −90° to 90° with a 1° step to record the angular scattering profile.

**Data availability**

The data that support the findings of study are available from the corresponding authors upon reasonable request.

**Acknowledgement**

The work at Zhejiang University was sponsored by the National Natural Science Foundation of China (NNSFC) (62222115, 62201500), the Key Research and Development Program of Zhejiang Province under Grant No.2024C01241(SD2), the Natural Science Foundation of Zhejiang Province under Grant No. ZCLY24F0101.

**Author contributions**

D.Z. and P.T. contributed equally to this work. Z.W., L.J., E.F. and H. C. initiated the plan and supervised the entire study. D.Z. and P.T. conceived the idea of this work and designed the simulations and experiments. D.Z., P.T., X.Y., B.Z., E.W. and Y. L. carried out the measurements and data analyses. D.Z., P.T. and Z. W. prepared the manuscript with input from all authors. All authors discussed the research.

**Competing interests**

The authors declare no competing interests.

**Additional information**

**Supplementary information.** The online version contains supplementary material available at XXX

**Correspondence** and requests for materials should be addressed to L. J., E.F., H.C., or Z.W.